\documentclass[amsmath,aps,prl,reprint,superscriptaddress]{revtex4-2}
\pdfoutput=1

\usepackage{bm}
\usepackage{graphicx}

\begin{document}

\title{Full-color photon-counting single-pixel imaging}

\author{Ya-Nan Zhao}
\affiliation{Hebei Key Laboratory of Optic-Electronic Information and
  Materials, College of Physics Science \& Technology, Hebei University,
  Baoding 071002, Hebei Province, China}

\author{Hong-Yun Hou}
\affiliation{Hebei Key Laboratory of Optic-Electronic Information and
  Materials, College of Physics Science \& Technology, Hebei University,
  Baoding 071002, Hebei Province, China}

\author{Jia-Cheng Han}
\affiliation{Hebei Key Laboratory of Optic-Electronic Information and
  Materials, College of Physics Science \& Technology, Hebei University,
  Baoding 071002, Hebei Province, China}

\author{Hong-Chao Liu}
\affiliation{Institute of Applied Physics and Materials Engineering,
  University of Macau, Avenida da Universidade, Taipa, Macao SAR, China}

\author{Su-Heng Zhang}
\email{shzhang@hbu.edu.cn}
\affiliation{Hebei Key Laboratory of Optic-Electronic Information and
  Materials, College of Physics Science \& Technology, Hebei University,
  Baoding 071002, Hebei Province, China}

\author{De-Zhong Cao}
\affiliation{School of Opto-Electric Information Science and Technology,
  Yantai University, Yantai 264005, Shandong Province, China}

\author{Bao-Lai Liang}
\email{liangbaolai@gmail.com}
\affiliation{Hebei Key Laboratory of Optic-Electronic Information and
  Materials, College of Physics Science \& Technology, Hebei University,
  Baoding 071002, Hebei Province, China}

\date{\today}

\begin{abstract}
We propose and experimentally demonstrate a high-efficiency single-pixel
imaging (SPI) scheme by integrating time-correlated single-photon
counting (TCSPC) with time-division multiplexing to acquire
full-color images at extremely low light level.
This SPI scheme uses a digital micromirror device to modulate a sequence
of laser pulses with preset delays to achieve three-color structured
illumination,
then employs a photomultiplier tube into the TCSPC module to achieve
photon-counting detection.
By exploiting the time-resolved capabilities of TCSPC,
we demodulate the spectrum-image-encoded signals, and then reconstruct
high-quality full-color images in a single-round of measurement.
Based on this scheme, the strategies such as single-step measurement,
high-speed projection, and undersampling
can further improve the imaging efficiency.
\end{abstract}

\maketitle

%%%%%%%%%%%%%%%%%%%%%%%%%%  body  %%%%%%%%%%%%%%%%%%%%%%%%%%
Single-pixel imaging (SPI), which reconstructs images from the signals of
a fixed single-pixel detector, has received increasing attentions in
recent years
\cite{gibsonSinglepixelImaging122020}.
As a representative of computational imaging schemes,
photon-counting single-pixel imaging (PCSPI) allows one to utilize
time-correlated single-photon counting (TCSPC) to capture intensity profiles
and three-dimensional structures at extremely low light level.
Benefiting from TCSPC, PCSPI has two significant performance advantages:
high sensitivity and precise timing resolution.
Combined with compressed sensing, spatial correlations, total variation
constraints, PCSPI enables fewer imaging photons
\cite{kirmaniFirstPhotonImaging2014,
  yangComputationalImagingBased2015,
  liuPhotonlimitedSinglepixelImaging2020},
faster imaging acquisition
\cite{howlandPhotonCountingCompressive2013,liuFastFirstphotonGhost2018},
longer imaging distances
\cite{pawlikowskaSinglephotonThreedimensionalImaging2017,
  liSinglephotonImaging2002021},
and lower imaging noise
\cite{chenRobustSinglephotonCounting2020}.
In addition, photon-counting detectors, especially photomultiplier
tubes (PMTs), have a wide spectral response
range. Therefore, the PCSPI should also enable full-color imaging
at extremely low light level.

In conventional SPI, a direct method to obtain full-color images is
to acquire the red (R), green (G), and blue (B) components separately
in three rounds of measurement.
However, this would consume triple of the time of monochromatic imaging,
making SPI inefficient and hard to be implemented.
Several methods have been proposed to enhance the imaging efficiency of
full-color SPI.
Welsh et al. proposed to use three spectrally filtered single-pixel detectors
to obtain the RGB components separately
\cite{welshFastFullcolorComputational2013}.
This approach can acquire full-color images in one round of measurement,
but increases the complexity of the detection system.
Subsequently, with the help of time-division multiplexing (TDM)
\cite{salvador-balaguerFullcolorStereoscopicImaging2015}
and spatial multi-modulation
\cite{huangMultispectralComputationalGhost2017,
  liEfficientSinglepixelMultispectral2017,
  zhangSimultaneousSpatialSpectral2018,
  yeGhostDifferenceImaging2021},
full-color SPI schemes with one single-pixel detector in one round of
measurement have been achieved.
These schemes can reduce the multispectral imaging time to
monochrome imaging time without significantly increasing the complexity of
the imaging system.
Recently, Kanno et al. used frequency-time-division multiplexing
to achieve ultrafast two-color single-pixel microscopy
\cite{kannoHighspeedSinglepixelImaging2020}.
In addition, by exploiting the sparsity of natural images and
human visual characters,
it has been found that appropriate undersampling can be performed
to further improve the efficiency of full-color SPI
\cite{yanColoredAdaptiveCompressed2016,
  qiuEfficientFullcolorSinglepixel2020}.
If these full-color SPI approaches are introduced
into photon-counting imaging to achieve efficient full-color PCSPI,
it will be of great interest on account of its important
applications under extreme environments, such as night vision,
remote sensing, and biological imaging.

In this letter, we propose and experimentally demonstrate a high-efficient
full-color PCSPI scheme.
We extend the spectral resolution capability of
PCSPI by integrating it with TDM.
Specifically, the proposed scheme utilizes a sequence of red, green, and blue
laser pulses with preset delays for light source,
a digital micromirror device (DMD) for structured illumination,
and a PMT with the help of TCSPC module for photon-counting detection.
By exploiting the powerful time-resolved capabilities of TCSPC,
we are able to demodulate spectrum-image-encoded signals, and then reconstruct
full-color images.
As a proof-of-principle demonstration, we use this method to experimentally
acquire a high-quality full-color image of a color slide of a small rocket
in one round of measurement through the Hadamard basis scan (HBS).
The strategies to further improve the imaging efficiency are also discussed,
such as using single-step measurement, increasing the projection rate,
and undersampling imaging.

\begin{figure}[tbp]
\centering
\includegraphics[width=0.995\columnwidth]{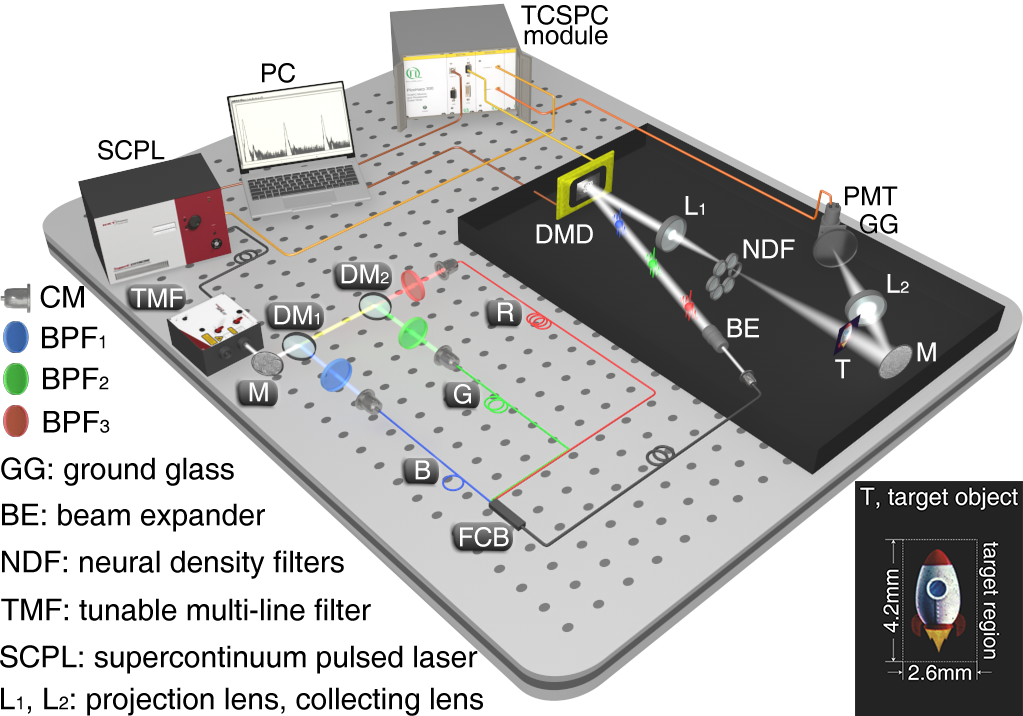}
\caption{Schematic of our full-color PCSPI.
  M, plane mirror;
  DM$_1$, DM$_2$, longpass dichroic mirrors with cut-on wavelengths
  of $ 505 $ nm and $ 605 $ nm;
  BPF$_1$, BPF$_2$, BPF$_3$, bandpass filters with $ 10 $ nm bandwidth
  and centered at $ 480 $ nm, $ 550 $ nm, and $ 670 $ nm;
  CM, multimode fiber collimator;
  FCB, $3\times1$ fiber combiner.}
\label{fig_scheme}
\end{figure}

The schematic of our proposed full-color PCSPI system is depicted
in Fig. \ref{fig_scheme}.
A supercontinuum picosecond pulsed laser (SuperK EXTREME, NKT) is
employed as the light source, and its repetition rate
is set at $ f_L = 15.6$ MHz.
A tunable multi-line filter (SuperK SELECT, NKT) is used to select
three monochromatic pulsed lights with center wavelengths of 480 nm, 550 nm,
and 670 nm from the supercontinuum laser and adjust their intensities
to the desired value (See Supplement 1, Section 1).
The selected three-wavelength pulsed beam is split into three delay paths by
two longpass dichroic mirrors (GCC-414003/6, Daheng Optics).
Three bandpass filters (BP480/550/670, Ruiyan) are placed
in three delay paths, respectively, to remove the light from other bands.
In each delay path, the monochromatic pulsed light is coupled into
a fiber of designed length to produce a suitable delay time.
Then, all three monochromatic pulsed lights with different pulses delay times
are combined into a single fiber via
a $3\times1$ fiber combiner (YS1N3, Yousheng).
This combined pulsed light is coupled into a beam expander (GCO-2503, Daheng)
via a collimator, and then incident on a DMD (F4100, Xintong).
The DMD is used to convert the incident pulsed beam into a sequence of
light patterns.
These light patterns work as structured illumination to be projected onto
the target object via a projection lens.
Neutral density filters are used to attenuate the light intensity
to the level of single-photon counting detection.
A color slide of a small rocket is used as the target object,
where the image is located within a rectangular region of
$ 2.6\text{ mm}\times4.2\text{ mm} $,
as shown by the inset of Fig. \ref{fig_scheme}.
The transmitted light from the color slide is redirected by a mirror to
a collecting lens, and then focused onto a PMT (R928, Hamamatsu)
that operates in Geiger-mode to detect arrivals of individual photon.
To improve the photon detection efficiency, a ground glass is placed
in front of the PMT to uniform the spatial intensity distribution of
the transmitted object light.
A TCSPC module (PicoHarp 300, PicoQuant) that is triggered by the periodic
sync signals from the pulsed laser, is used to record
the photon count events from PMT and the marker signals from DMD.
The TCSPC module works in time-tagged time-resolved (TTTR) mode so that
the delay time of each photon event with respect to
the previous laser sync event is recorded with picosecond resolution.
In addition, the photon events and the marker events are time tagged
by counting sync events from the beginning of the experiment .

To efficiently obtain full-color imaging, the approach of HBS is used.
The two-dimensional Hadamard transform pair is defined as
\begin{align}
  \bm{F}&=\bm{H}\bm{f}\bm{H},\\
  \bm{f}&=\frac{1}{N^2}
  \bm{H}\bm{F}\bm{H}\label{eq_InverseHT},
\end{align}
where $ \bm{f} $ represents the spatial sampling matrix of the object,
$ \bm{F} $ represents the spectrum matrix of the object,
$\bm{H}$ denotes the Walsh ordered Hadamard matrix,
$ N $ is the order of Hadamard matrix.
It is worth noting that $ \bm{f} $ needs to be a square matrix on
the order of $ N $.
The complete set of Hadamard basis patterns can be generated as follows
\begin{equation}
  \left\{\bm{P}_{uv}|\bm{P}_{uv}=
    \bm{H}^T_u\bm{H}_v,\;u,\,v=0,1,\cdots,N-1
  \right\},
\end{equation}
where $\bm{H}_u$, $\bm{H}_v$ are the $ u $-th, $ v $-th
row vectors of $\bm{H}$, respectively.
Then the spectrum coefficient $ F(u,v) $ can be written as
\begin{equation}\label{eq_basiscan}
  F(u,v)=\langle\bm{f}, \bm{P}_{uv}\rangle_{\mathrm{F}},
  \quad u,v=0,1,\cdots,N-1,
\end{equation}
where the $ \langle\cdot,\cdot\rangle_{\mathrm{F}} $ denotes the Frobenius
inner product.
According to Eq. (\ref{eq_basiscan}), each spectrum coefficient
can be acquired one by one, hence this method is called HBS.

Differential measurement is a common implementation of HBS.
In order to acquire each spectrum coefficient $ F(u,v) $ using
intensity modulation, each basis pattern $ \bm{P}_{uv} $ is
transformed into two non-negative patterns as follows
\begin{align}
  \bm{P}^{(+)}_{uv}&
  =\frac{1}{2}\left(1+\bm{P}_{uv}\right),\label{eq_PosPattern}\\
  \bm{P}^{(-)}_{uv}&
  =\frac{1}{2}\left(1-\bm{P}_{uv}\right).
\end{align}
These two non-negative patterns are projected onto the object in turn, and the
transmitted object light intensities measured by the single-pixel detector are
\begin{align}
  D^{(+)}(u,v)&=k\langle\bm{f}, I_s\bm{P}^{(+)}_{uv}
    +I_ab(u,v)\rangle_{\mathrm{F}},\label{eq_PosPro}\\
  D^{(-)}(u,v)&=k\langle\bm{f}, I_s\bm{P}^{(-)}_{uv}
    +I_ab(u,v)\rangle_{\mathrm{F}},
\end{align}
where $ k $ represents the quantum efficiency of the single-pixel detector,
$ I_s $ represents the light source intensity, and $ I_a b(u,v) $ represents
background light intensity.
By differential intensity signals between the positive and inverse patterns,
the corresponding spectrum coefficient can be achieved, i.e.
\begin{equation}
  F(u,v)=\alpha[D^{(+)}(u,v)-D^{(-)}(u,v)],
\end{equation}
where $ \alpha=1/kI_s $.
The background light hardly changes during the measurement of
the couple of patterns.
Therefore, the differential method can also remove the background light
very well.
During the imaging process, the whole set of non-negative patterns
\begin{equation}
  \left\{\bm{P}^{(+)}_{uv},\;
    \bm{P}^{(-)}_{uv}\big|u,\,v=0,1,\cdots,N-1
  \right\}
\end{equation}
are projected one by one, and then the spectrum of the object can be
acquired as
\begin{equation}\label{eq_HSpec}
  \bm{F}=\alpha\left[\bm{D}^{(+)}-\bm{D}^{(-)}\right].
\end{equation}
Finally, according to Eq. (\ref{eq_InverseHT}), the image of the object can
be reconstructed by inverting the Hadamard spectrum.

Differential measurement requires $ 2N^2 $ non-negative basis
patterns to reconstruct an $ N\times N $ pixels image.
In present experiments the resolution of all reconstructed images
is set to $ 64\times64$ pixels.
Thus, for differential measurement, the total number of non-negative patterns
to be projected is $ M=8192$.
The projection rate of DMD is set at $ f_D=1000 $ Hz.
Therefore, the image acquisition time is $ t_A=M/f_D=8.192 $ s.
During image acquisition, the light pulses output by the laser
are $ N_L=f_L\times t_A\times 3\approx3.834\times 10^8 $,
in which the factor of 3 comes from RGB light pulses,
and the total photon counts given by the PMT are about $7.644\times 10^6$.
Thus, only {2\%} of laser pulses generate a photon count at the PMT,
which maintains single-photon counting detection.

\begin{figure}[htbp]
\centering
\includegraphics[width=0.95\columnwidth]{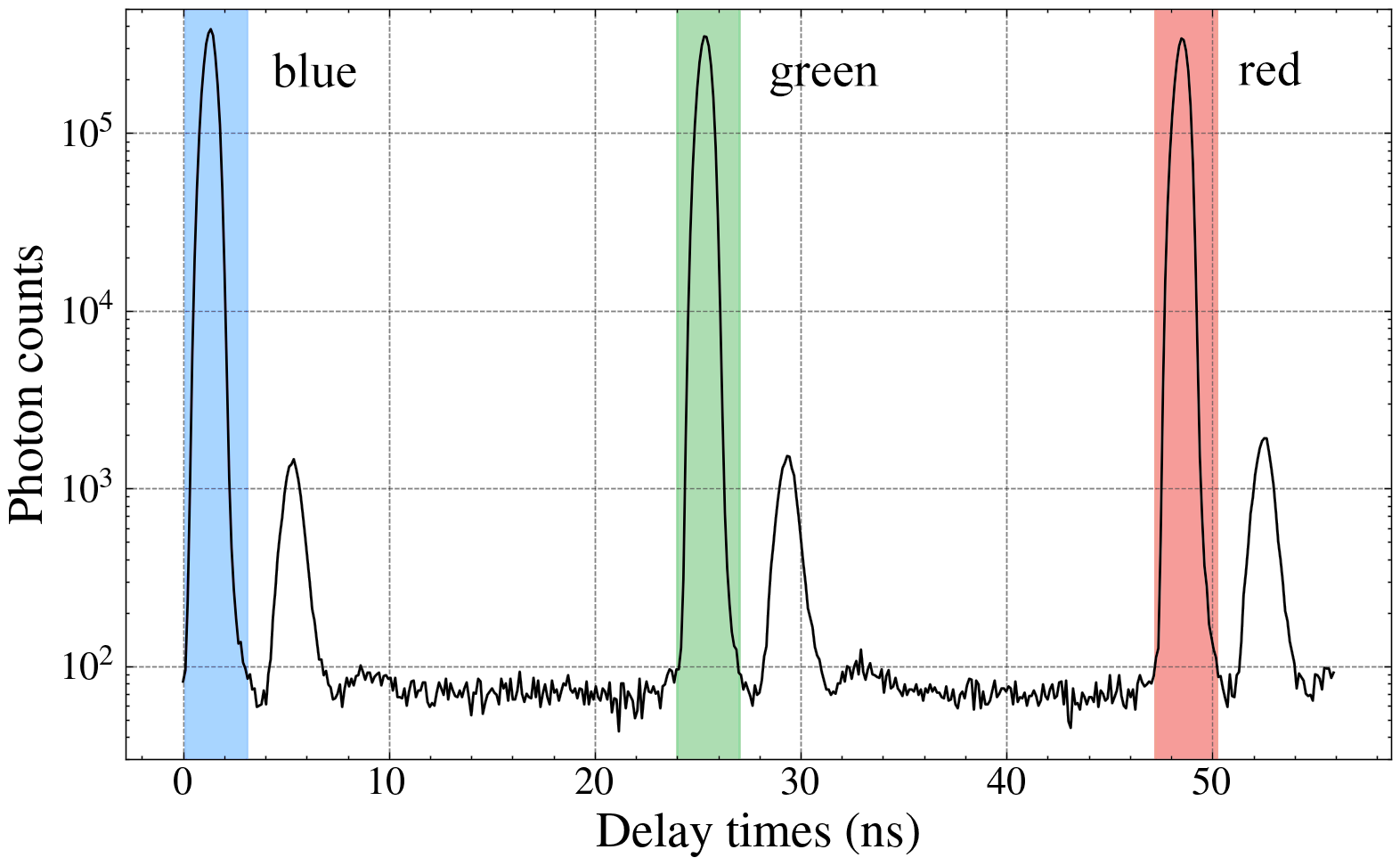}
\caption{The histogram of the photon counts with respect to the delay times
during the image acquisition.
The width of the delay time bin is $ 16 $ ps.
The vertical axis is log-10 scale.
The photon counts located in the B, G and R shaded regions are
generated by the transmitted object light of the corresponding color.
The three associated peaks are formed due to afterpulses.}
\label{fig_phist}
\end{figure}

The TCSPC module records the delay time for each photon event with respect to
the sync laser trigger.
So a histogram of the photon counts can be drawn with respect to
the delay time, as shown in Fig. \ref{fig_phist}.
There are three main peaks, indicated by RGB ribbons, respectively.
They are generated by the transmitted object light pulses of the
corresponding color.
The width of these main peaks is $3$ ns, depending on the response time
of the TCSPC system, and the separation between main peaks is about
$24$ ns, depending on the optical path difference between
the delay paths.
Three associated peaks are also observed, but these are fake signals
formed due to the afterpulses in the PMT.

\begin{figure}[htbp]
\centering
\includegraphics[width=\columnwidth]{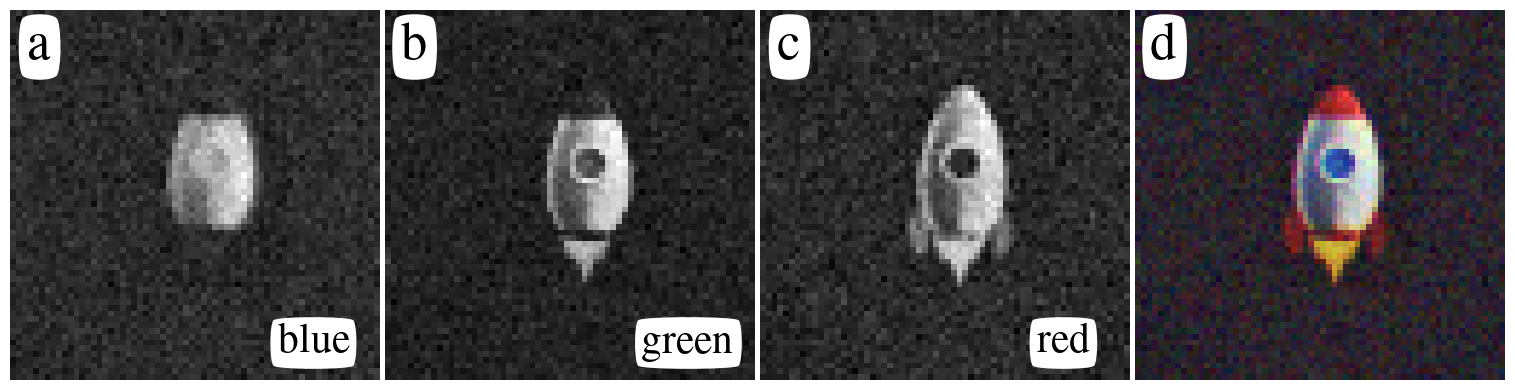}
\caption{Experimentally reconstructed images with differential measurement.
  (a) -- (c) B, G , and R component images reconstructed from the
  corresponding photon counts;
  (d) full-color composite image.}
\label{fig_recimag}
\end{figure}

The photon counts within each main peak are highly time-correlated,
being overwhelmingly generated by the signal light pulses of the
corresponding color,
while the photon counts outside the peak are mainly background counts
and dark counts.
In order to reduce the impacts from dark counts and background counts,
only photon counts within the three shaded regions are selected to
reconstruct the corresponding RGB color component of the full-color image.
Based on the time-tag of the photon events, each photon count can correlate
with the projection pattern that generated it.
Then all the photon counts from the same projection pattern are
accumulated as the bucket intensity measured by the PMT.
By using Eqs. (\ref{eq_HSpec}) and (\ref{eq_InverseHT}), the B, G,
and R component images can be reconstructed from the corresponding
photon counts, as shown in Figs. \ref{fig_recimag}(a) -- (c).
The ultimate full-color image, as shown in Fig. \ref{fig_recimag}(d),
has good imaging quality with normal color.
It should be noted that the relative height of the main peaks
in the photon counts histogram affect the color balance of
the reconstructed ultimate full-color image.
In our experiments, we made the heights of the three main peaks to
be nearly equal, as shown in Fig. \ref{fig_phist},
through independent adjustment of each color light intensity,
and then the ultimate full-color image showed a good color balance.
However, if the heights of the main peaks are significantly different,
the ultimate full-color image will show color imbalance
(See Supplement 1, Section 2).

In order to obtain appropriate spatial resolution and imaging range,
the projection patterns are resized to $ 192 \times 192 $ pixels by the
nearest-neighbor interpolation process and then displayed onto the DMD.
The light patterns modulated by the DMD are projected onto
the object at a magnification of $ 2.52 $.
Thus the spatial resolution and size of the reconstructed image are
about $ 103.42 $ {\textmu}m and $ 6.62 $ mm, respectively.

The above differential method requires two measurements to obtain a
spectrum coefficient, which can effectively eliminate the influence
of background light, but doubles measurements.
Recently, Xiao et al. proposed a single-step measurement method
to reduce half of the measurements
\cite{xiaoDirectSingleStepMeasurement2019}.
Its principle is as follows.
By substituting Eq. (\ref{eq_PosPattern}) into Eq. (\ref{eq_PosPro}) yields
\begin{equation}
  D^{(+)}(u,v)=\frac{kI_s}{2}F(0,0)+\frac{kI_s}{2}F(u,v)+kI_aF(0,0)b(u,v).
\end{equation}
Let $ F'(u,v)=\alpha[2D^{(+)}(u,v)-D^{(+)}(0,0)] $,
then it can be find that $ F'(u,v)=F(u,v)+\beta b'(u,v) $,
where $ \beta=(I_a/I_s)F(0,0) $, $ b'(u,v)=2b(u,v)-b(0,0) $.
This indicates that each spectrum coefficient can be obtained
by just a single-step measurement.
During image acquisition, only the positive patterns are measured,
and the entire Hadamard spectrum can be obtained as
\begin{equation}
  \bm{F}'=\alpha[2\bm{D}^{(+)}-D^{(+)}(0,0)].
\end{equation}
However, the measured spectrum contains a background term, i.e.
$ \bm{F}'=\bm{F}+\beta \bm{b}' $.
If the Hadamard inversion of the measured spectrum is performed directly,
the reconstructed image can be expressed as
$ \bm{f}'=\bm{f}+\beta\bm{B}'/N^2 $,
where $ \bm{B}'=\bm{B}+\bm{B}_e $,
$ \bm{B}=\bm{H}\bm{b}\bm{H} $,
$ \bm{B}_e=\bm{H}\left[\bm{b}-b(0,0)\right]\bm{H} $.
Note that the influence of the background light on each measured spectrum
coefficient is statistically homogeneous, but not on the reconstructed image.

The energy of the background spectrum $ \bm{B} $ is mainly concentrated
around the upper left corner,
while the relative changes $ \bm{b}-b(0,0) $ of the background
are associated with the high-frequency components,
and the energy of its spectrum $ \bm{B}_e $ is mainly concentrated
around the other corners.
Therefore, when the image of the object has no energy distribution at the edge
of the imaging area, the background term can be separated for
reasonable estimation.
Let $ \beta \bm{\hat{B}}'/N^2 $ represents the estimation of the
background term, the estimation of the Hadamard spectrum of object can
be written as $ \bm{\hat{F}}=\bm{F}'-\beta \bm{\hat{b}}' $, where
$\bm{\hat{b}}'=\bm{H}\bm{\hat{B}}'\bm{H}/N^2$.
Then perform the inverse Hadamard transform to obtain the estimation of the
object image.

\begin{figure}[htbp]
\centering
\includegraphics[width=\columnwidth]{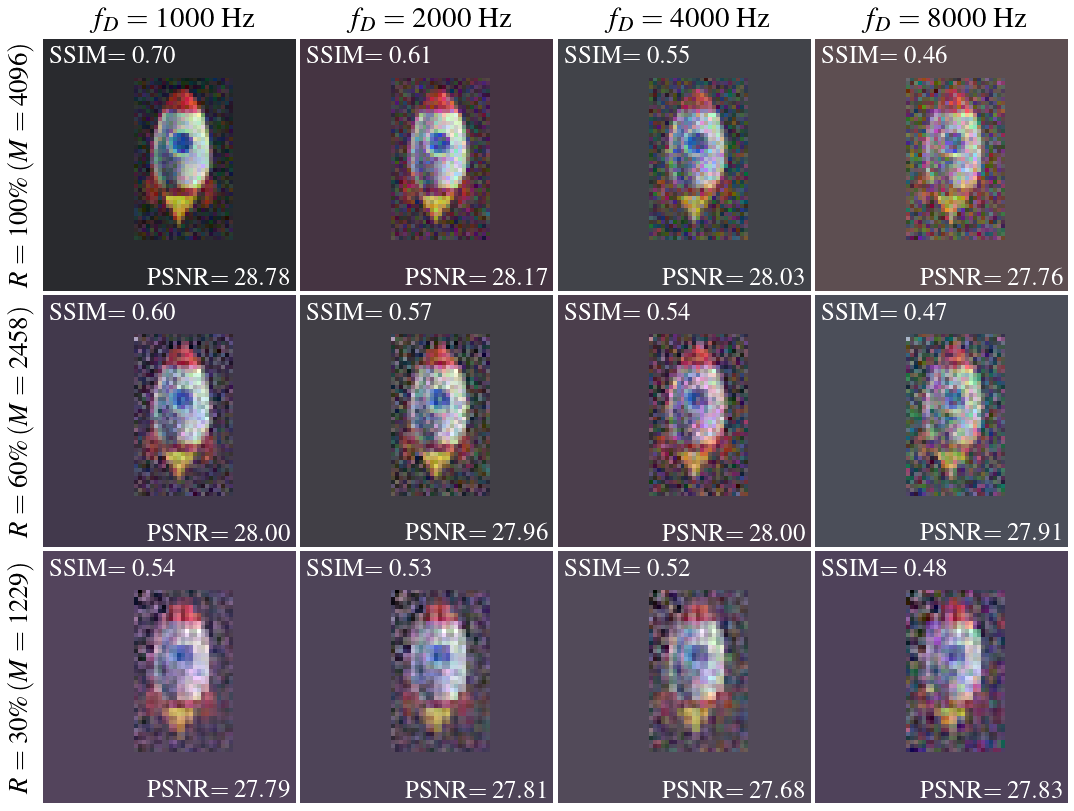}
\caption{Experimentally reconstructed images with single-step measurement
  at different projection rates and different sampling ratios.}
\label{fig_singlestep}
\end{figure}

Based on the single-step measurement technique, one can increase
the projection rate of DMD, and reduce the measurements by under-sampling
to further improve efficiency.
The reconstructed images using the single-step method
with different projection rates and different sampling ratios
are shown in Fig. \ref{fig_singlestep}.
Since the rocket image is located within the target region, the part outside
the target region in the reconstructed image is considered as
the background term (See Supplement 1, section 3).
As the projection rate $ f_D $ increases, the noise of the reconstructed
image increases significantly.
This is because the projection rate $ f_D $ increases, the dwell time of
every single projection pattern is shortened,
and the photon counts are more affected by the Poisson noise.
As the sampling ratio $ R $ decreases, the reconstructed image
gradually lost its details.
However, the noise is also reduced by spatially smoothing.
Therefore, it is possible to quickly acquire the contours of the object at
high projection rate and low sampling ratio.

For reference,  the structural similarity index (SSIM) and the peak
signal-to-noise ratio (PSNR) are used to quantitatively evaluate the
quality of reconstructed images in Fig. \ref{fig_singlestep}.
The calculation of these quality parameters is based on the
target region of the differential reconstructed image.
The reconstructed image quality with single-step
measurement is more susceptible to the sampling strategy.
Although the commonly used zig-zag strategy can preserve more details
\cite{zhangHadamardSinglepixelImaging2017},
it may also introduce more noise.
To solve this issue, we propose a new sampling strategy,
named hyperbolic sampling strategy,
which only acquire the spectrum coefficients whose spatial
frequency $ u,\,v $ satisfy
$ \left[(N-1-u)(N-1-v)\right]^{{1}/{2}}\ge\rho_c $,
where $ \rho_c $ is called cutoff frequency, and takes the values in the
range $ [0,\;N-1] $ (See Supplement 1, Section 4).

In conclusion, we demonstrate an efficient full-color PCSPI scheme.
This PCSPI scheme fully exploits the time-resolved capabilities of
TCSPC to acquire a full-color image, using only one single-pixel detector
in a single round of measurements.
Benefiting from TDM, this method takes time to acquire a full-color
image the same as that for a monochromatic image.
Based on this scheme, one can also use single-step measurement, high-speed
projection, and undersampling to further improve imaging efficiency,
which offers a possibility to achieve full-color imaging of 3D realistic
moving objects at extremely low light level.
In addition, this scheme can also be adopted by other computational imaging
techniques to achieve multispectral imaging.

\section*{Funding}
Natural Science Foundation of Hebei province (F2019201446);
Start-up Research Grant of University of Macau (SRG2019-00174-IAPME);
National Natural Science Foundation of China (NSFC) (11204062, 11674273).

\section*{Disclosures}
The authors declare no conflicts of interest.

\section*{Supplement document}
See Supplement 1 for supporting content.

% \bibliographystyle{apsrev4-2}
\bibliography{Reference}

\end{document}